# An insight into the importance of national university rankings in an international context: The case of the I-UGR Rankings of Spanish universities[1]

Nicolás Robinson-García[1], Daniel Torres-Salinas[2]*, Emilio Delgado López-Cózar[1] and Francisco Herrera[3]

[1] *{elrobin, edelgado} @ugr.es*
EC3: Evaluación de la Ciencia y de la Comunicación Científica, Departamento de Biblioteconomía y Documentación, Universidad de Granada (Spain)

[2] *torressalinas@gmail.com*
EC3: Evaluación de la Ciencia y de la Comunicación Científica, Centro de Investigación Biomédica Aplicada, Universidad de Navarra (Spain)

[3] *herrera@decsai.ugr.es*
Department of Computer Science and Artificial Intelligence, Universidad de Granada (Spain)

* To whom all correspondence should be addressed

**Abstract**
The great importance international rankings have achieved in the research policy arena warns against many threats consequence of the flaws and shortcomings these tools present. One of them has to do with the inability to accurately represent national university systems as their original purpose is only to rank world-class universities. Another one has to do with the lack of representativeness of universities' disciplinary profiles as they usually provide a unique table. Although some rankings offer a great coverage and others offer league tables by fields, no international ranking does both. In order to surpass such limitation from a research policy viewpoint, this paper analyzes the possibility of using national rankings in order to complement international rankings. For this, we analyze the Spanish university system as a study case presenting the I-UGR Rankings for Spanish universities by fields and subfields. Then, we compare their results with those obtained by the Shanghai Ranking, the QS Ranking, the Leiden Ranking and the NTU Ranking, as they all have basic common grounds which allow such comparison. We conclude that it is advisable to use national rankings in order to complement international rankings, however we observe that this must be done with certain caution as they differ on the methodology employed as well as on the construction of the fields.

## 1. Introduction

Since the launch of the first edition of the Shanghai Ranking in 2003, interest has grown on the development of tools for benchmarking and comparing academic and research institutions. As a result of the massification of higher education, the race for excellence and a fierce battle for research funding, universities now strive for positioning themselves in these international rankings (Hazelkorn 2011). These tools have gained an undisputable position in research managers' 'toolkit' for measuring the state of health of higher education institutions and the main resource for many universities and countries when taking decisions in a research policy context (Marginson & van der Wende, 2007). The great effect they have- not only in the media and the public but also for research managers, politicians and decision makers - relies on the perception that highly ranked institutions are usually more productive, produce higher quality research and teaching and contribute best to society than the rest of universities (Shin & Toutkoushian, 2011).

---

[1]This is a revised and updated version of a communication presented at the ISSI Conference 2013 held in Vienna

However, despite their advantages as easy-to-read tools, they also have many inconsistencies and shortcomings that warn against a careless use (Delgado López-Cózar, 2012). In this sense, we can identify five major issues which must be addressed: 1) methodological and technical errors and difficulties such as the recollection of reliable and standardized data (Toutkoushian & Webber, 2011); 2) the criteria for selecting the indicators are not scientifically supported (Van Raan, 2005); 3) the multidimensional nature of universities (Orduña-Malea, 2012; Waltman et al., 2012) leads to a wide heterogeneity among institutions (Collini 2011); 4) using a unique table to rank universities neglects their disciplinary focus (Visser et al., 2007); and 5) international rankings cannot reflect the state of national higher education systems as they usually cover just the top universities of each country (Torres-Salinas et al., 2011a).

While the issue of data reliability still remains a major shortcoming and there is no consensus yet on which indicators represent better the nature and quality of universities, the other issues have been somehow surpassed using approaches which do not solve completely their dangers but, at least, diminish the flaws. For instance, rankings such as the Leiden Ranking (Waltman et al., 2012) or the Scimago Institutions Rankings (henceforth SIR) have emerged focusing uniquely on the research dimension of universities to the neglect of other aspects such as innovation or teaching. Others now publish, along with a global ranking, rankings by subjects and fields, which offer a better picture of universities' performance (García et al, 2012). Also, some rankings such as the SIR or the Ranking Web of World Universities cover now not just top-class universities but the former includes more than 3,000 research institutions and the latter, more than 19,000.

Rankings have not been fully developed and still have serious shortcomings (van Raan, 2005). But their dominance as decisive factors in research policy (Hazelkorn, 2011) at national and supranational level puts them in the spotlight. One of the most important threats rankings entail is that they ignore universities' diversity, which can affect seriously the health of higher education systems and lead to dangerous and simplistic conclusions when interpreting and developing ranking systems (e.g., Moed et al., 2011). These differences affect institutions at two levels, at their organizational structure, and in the national configuration of higher education systems, affecting their multidisciplinary nature and diversity (Orduña, 2011). The phenomenon of university rankings has influenced deeply all university systems, even those that were not conceived at first to establish a competitive framework. Therefore, in order to analyze the success or failure of different countries in their research policy, university systems should be assessed as a whole, and not considering each university as an individual and autonomous unit. Such approach was applied by Docampo (2011) using the Shanghai Ranking in order to analyze the university systems of the countries represented.

Despite its limitations, this study offers a glimpse of the global scenario regarding the research excellence of different countries' university systems. In Table 1 we show the clusters emerged from the study carried out by Docampo (2011) and the number of universities by country in different intervals according to the 2012 edition of the Shanghai Ranking. Therefore we observe a dominance of the United States and the United Kingdom which alone represent more than a third of the universities included in the ranking (37.6%), followed by Germany and Canada as the next with the highest number of universities included. However, despite the numbers, except Japan, which in this new edition includes a university in the top20, none of the others have a university positioned within this interval.In this context, the truth is that the high visibility Anglo-Saxon universities have in rankings leaves little space for others, blurring the state of other countries which are working towards a successful university model. In fact, it clearly shows the incapability of the ranking to represent national university systems with exhaustiveness.

**Table 1.** University systems by country considering the results in Docampo (2011) and the 2012 Shanghai Ranking edition. Leaders, Fast followers and followers

|  | Countries | Nr of Universities Top20 | Nr of Universities Top100 | Nr of Universities Top300 | Nr of Universities Top500 |
|---|---|---|---|---|---|
| Leaders | United States | 17 | 53 | 109 | 150 |
|  | United Kingdom | 2 | 9 | 30 | 38 |
|  | Switzerland | --- | 4 | 7 | 7 |
| Fast followers | Australia | --- | 5 | 9 | 19 |
|  | Canada | --- | 4 | 17 | 22 |
|  | Sweden | --- | 3 | 7 | 11 |
|  | Israel | --- | 3 | 4 | 6 |
|  | Netherlands | --- | 2 | 10 | 13 |
|  | Denmark | --- | 2 | 4 | 4 |
| Followers | Germany | --- | 4 | 24 | 37 |
|  | France | --- | 3 | 13 | 20 |
|  | Belgium | --- | 1 | 6 | 7 |
|  | Norway | --- | 1 | 3 | 4 |
|  | Finland | --- | 1 | 1 | 5 |

Thus, these rankings do not offer a complete view of national higher education systems, preventing research managers and decision makers to have an accurate picture of the state of each country's university system. Hence the need for developing tools with higher levels of granularity in the information provided by rankings (Bornmann, Mutz & Daniel, 2013). For this reason, in 2010 members from the EC3 and Soft Computing research groups developed the Rankings I-UGR of Spanish Universities according to Fields and Scientific Disciplines (henceforth I-UGR Rankings) available at http://rankinguniversidades.es. It was originally named ISI Rankings but changed to its current name in its 2012 edition. This website offers 49 rankings for Spanish universities divided in 12 fields and 37 disciplines, according to their international research performance. Spain is a good example of a misrepresented higher education system. For instance, in the 2013 edition of the Shanghai Ranking only 10 universities out of 74 met the criteria for inclusion in the global ranking. In fact, none made it to the top 100 and only four were included in the 201-300 interval. Also, as it occurs with other countries such as Italy (Abramo, Cicero & D'Angelo, 2011), it is a non-competitive higher education system, which means that universities do not act as individual units but within a national framework, therefore decisions should not be taken relying on such a poor sample.

The main goal of the present paper is analyze if national rankings are necessary complements to international rankings. This paper is focused at the potential use of the information provided by national and international rankings by research managers and intends to explore if the information provided by both types of rankings is complementary and useful from a research management perspective. For this we will use the I-UGR Rankings, in order to:

    1) Analyze if national ranking are necessary complementing the information provided by international rankings, as the latter do not represent well national university systems.

    2) Analyze the levels of agreement between national and international rankings regarding the following aspects:

a. Are the top Spanish universities the ones visible in international rankings?

b. Disciplinary concordance: Do the different classifications by fields and subjects allow an analysis by areas?

To develop this study we select the Shanghai Ranking, the Times Higher Education World University Ranking (henceforth THE Ranking), the QS Ranking, the Leiden Ranking and the National Taiwan University Ranking (henceforth NTU Ranking). The first one to include disciplinary-oriented league tables was the Shanghai Ranking, launching in 2007 rankings by five broad fields and in 2009 five more rankings in specific disciplines, followed by the THE Ranking, the QS Rankings and the NTU Ranking. The Leiden Ranking has been the last one to follow this trend and now includes in its last edition rankings by five broad areas.

The paper is structured as follows. In section 2 we present the Spanish case analyzing its current state and we introduce the I-UGR Rankings, we contextualize its creation and we describe the methodology employed for their development. In section 3 we address the main issue of this paper: we compare the results of the main international rankings and the I-UGR Rankings for Spanish universities. Finally, in Section 4 we resume our main findings and their consequences in a research policy scenario.

**2. Spain as a case study: introduction to the I-UGR Rankings**

The Spanish university system is formed by 74 universities: 48 public and 26 private. However in the 2013 edition of the Shanghai Ranking only 10 met the minimum requirements to be included. It is a country poorly represented in the main international rankings due to the scarce number of universities considered as World-Class universities. But the impact these rankings have in research policy threatens a good governance and sensible decision making as they do not offer a complete picture of the university system (Docampo, 2011). In fact, as observed in Table 2, only 19 universities (18 public and 1 private universities) are included in four of the most important rankings; that is, 25.68% of the whole system. For this reason, other tools are needed in order to complete this fragmented picture of the Spanish higher education scenario.

**Table 2.** Spanish universities represented within the top 500 universities in the 2013 edition of the Shanghai Ranking, the QS Ranking, the Leiden Ranking and the NTU Ranking

| Shanghai Ranking | | Leiden Ranking | | QS Ranking | | NTU Ranking | |
|---|---|---|---|---|---|---|---|
| Barcelona | 201-300 | Barcelona | 259 | Aut Barcelona | 177 | Barcelona | 89 |
| Aut Madrid | 201-300 | Pol Valencia | 282 | Barcelona | 178 | Aut Barcelona | 169 |
| Aut Barcelona | 201-300 | Santiago | 317 | Aut Madrid | 195 | Aut Madrid | 214 |
| Complutense | 201-400 | Aut Barcelona | 333 | Complutense | 216 | Valencia | 224 |
| Pol Valencia | 301-400 | Valencia | 336 | Pompeu Fabra | 281 | Complutense | 259 |
| Valencia | 301-400 | Aut Madrid | 356 | Navarra | 315 | Granada | 267 |
| Pompeu Fabra | 301-400 | Zaragoza | 366 | Carlos III | 317 | Oviedo | 369 |
| Granada | 301-400 | Granada | 375 | Pol Cataluña | 345 | Santiago | 378 |
| Zaragoza | 401-500 | Pol Cataluña | 396 | Pol Valencia | 383 | Zaragoza | 392 |
| País Vasco | 401-500 | Sevilla | 402 | Pol Madrid | 389 | País Vasco | 421 |
| | | Complutense | 406 | Salamanca | 441-450 | Sevilla | 434 |
| | | Murcia | 408 | Valencia | 471-480 | Pol Valencia | 446 |
| | | Oviedo | 409 | Zaragoza | 481-490 | Pompeu Fabra | 463 |
| | | País Vasco | 411 | | | | |
| | | Pol Madrid | 434 | | | | |

**List of abbreviations used:** Aut: Autónoma; Pol: Politécnica

The first edition of the I-UGR Rankings was launched in 2010. Its development was motivated by the scarce visibility Spanish universities have in international rankings, which leads to a fragmented picture of the Spanish university system. Though other national rankings had already been developed, these were considered insufficient due to the limitations they presented which made them unsuitable as research policy tools. Among other limitations we address the following: lack of continuity over time, exclusion of private institutions, disregard of disciplinary focus, use of rudimentary bibliometric indicators, selection of unsuitable time periods or election of databases with dubious selection criteria of sources (Torres-Salinas et al., 2011a).

Data is retrieved from the Thomson Reuters Science Citation Index and Social Science Citation Index (SCI and SSCI).The reason for using such source database relies not only on its importance as a bibliometric database containing the main international scientific literature, but also due to its importance in the Spanish research evaluation system (Cabezas-Clavijo et al., 2013). In its first edition 12 rankings were offered for 12 broad fields. These fields were later expanded with 19 subfields or disciplines in the second edition (Torres-Salinas et al., 2011b) and finally, 37 disciplines in the 2012 edition. The fields and disciplines were constructed by aggregating the subject categories to which records from the Journal Citation Reports are assigned. Aggregating subject categories is a classical perspective followed in many bibliometric studies when adopting a macro-level approach (e.g., Moed, 2005; Leydesdorff & Rafols, 2009). For further information on the coverage of the I-UGR Rankings and the development of the fields and subfields the reader is referred to the following document in which methodology of the indicator for ranking universities as well as the construction of fields are defined[2].

Once the data is compiled into a relational database, the indicators defined in Table 3 are computed, and the index for rating each university is calculated. To rank universities we use the IFQ$^2$A Index (Torres-Salinas et al., 2011c). This indicator measures the quantitative and qualitative dimensions of the research outcome of a group of institutions in a given field. It is based on six primary bibliometric indicators, three focused on the quantitative dimension (QNIF) and the other three focused on the qualitative dimension (QLIF). These two dimensions represent two different aspects of the research activity, impact and visibility of universities. While the QNIF is based on size-dependent measures, the QLIF relies on relative measures of impact (as defined by the citations received) and visibility (as defined by the quartile to which a journal belongs according to its Impact Factor and the top papers among the 10% most cited papers). QLIF is a no size-dependent measure. In Table 3 we summarize the methodology employed for calculating the IFQ$^2$A Index. More information about the IFQ$^2$A Index may be found in Torres-Salinas et al. (2011c).

---

[2] http://sci2s.ugr.es/rankinguniversidades/downloads/rankingsI-UGR_Methodology_EV.pdf

**Table 3.** Calculation of the IFQ2A Index and definition of indicators.

| $QNIF = \sqrt[3]{NDOC \times NCIT \times H}$ | | $QLIF = \sqrt[3]{\%1Q \times ACIT \times TOPCIT}$ | |
|---|---|---|---|
| NDOC | Number of citable papers published in scientific journals | %1Q | Ratio of papers published in journals in the top JCR quartile |
| NCIT | Number of citations received by all citable papers | ACIT | Average number of citations received by all citable papers |
| H | H-Index as proposed by Hirsch (2005), over all the publications of the institution | TOPCIT | Ratio of papers belonging to the top 10% most cited papers calculated within all institutions |
| $IFQ^2A = QNIF \times QLIF$ | | | |

The selection of the indicators as well as the conceptualization of the index, are based on the following criteria:

1) The indicators chosen must not be restrictive. That is, they should be applied to all institutions. For instance, the Shanghai Ranking uses the number of Nobel Prizes as an indicator to measure research excellence. In the Spanish case only one university is affected by it (Complutense de Madrid).

2) Rankings must be size-independent, however if the numbers are too small they may distort the ranking and introduce a certain degree of instability. This leads to the use of a bidimensional index which takes into account raw counts of papers and citations as well as relative measures which benefit small institutions which produce high quality papers (as defined by bibliometric indicators).

3) Rankings must take into account the disciplinary focus of universities. For this, a unique list cannot be provided. Contrarily, one must offer rankings by field of specialization in order to provide useful tools for research managers.

4) Seniority must not be rewarded. For this, fixed time periods must be used. Also, when calculating the H-Index, this must be considering the time frame used. In this sense, the I-UGR Rankings offer a five-year window and a ten-year window.

5) Stability must be assured. This means that the fixed time frame must be wide enough to offer stable results. A five-year time frame allows results to be consistent and significant.

**Figure 1.** Distribution of universities according to their qualitative and quantitative dimensions in the field of Computer Science. 2008-2012. Top 5 institutions according to the IFQ$^2$A Index are highlighted and labeled.

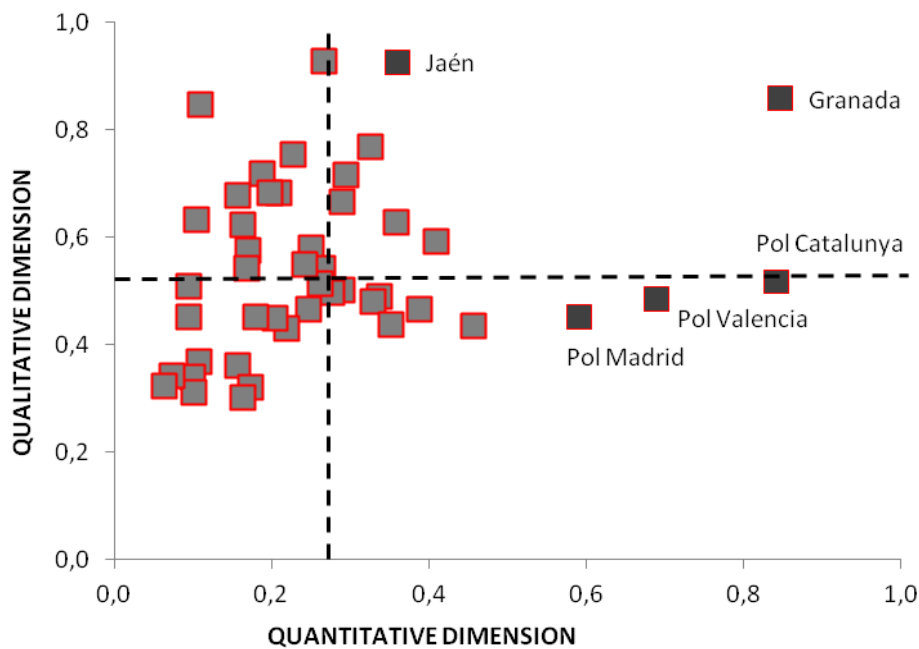

In Figure 1 we show the distribution of universities according to the QNIF and QLIF in the field of Computer Science for the 2008-2012 time period. The dashed lines show the average values of each dimension. Universities positioned at the top right hand of the figure are those which outstand in both dimensions. Those positioned on the bottom right outstand on the quantitative dimension but not on the qualitative dimension. At the top left, we observe university with small research output but high quality research. Lastly, in the bottom left, universities which do not outstand in any dimension are represented. As we can observe, although top universities outstand in both dimensions, many universities outstand in the qualitative dimension but do not do so in the quantitative dimension. Due to the bidimensional nature of the IFQ$^2$A index, these small institutions are reflected in the rankings.

## 3. Levels of agreement and disciplinary concordance between rankings: Comparison by fields of the main international rankings and the I-UGR rankings

In this section we analyze the state of the Spanish university system using international and national rankings. For this, we first establish in Section 3.1 a set of criteria for the selection of the rankings we will use in order to set some basic common grounds which will allow a fair comparison between them. Then, in Section 3.2 we match rankings by fields between the international and national rankings and finally, we analyze the level of agreement between them. For this we use two indicators. On the one hand, we calculate the Spearman's rank correlation coefficient or Spearman's rho, which will indicate to what extent are the different rankings coherent between them on the order in which Spanish universities are displayed. On the other hand, we show the level of agreement between rankings, which indicates if universities included in an international ranking coincide with those which occupy the top positions of the national ranking.

*3.1 Selection of international rankings*

The aim is to use international and national rankings as complementary tools to offer on the one hand, a global perspective of the position of Spanish universities and, on the other hand, a complete picture of the Spanish university system. For this, we first need to establish a set of criteria for choosing the most relevant rankings for our purposes. These are the following:

> 1) Rankings must be based on the research performance of universities, at least partially, as we are analyzing the research dimension of universities.
>
> 2) Data retrieved for the construction of the rankings must come from a reliable bibliometric database or information resource, at least partially.
>
> 3) They must offer rankings by fields, as we have considered that only this way we can provide an accurate image of universities' research performance.

Based on these criteria we selected the I-UGR Rankings as national rankings and the following international rankings. In table 4 we include the main characteristics of each of these rankings. For more detailed information on the methodology of each ranking, we refer the reader to its website; here we will briefly describe them:

> *1) Shanghai Ranking* (http://www.shanghairanking.com/). It was not only the first international ranking launched (Liu & Cheng, 2005) but it is used as yardstick to measure the research excellence of universities worldwide (Docampo, 2011). It is based on six indicators, two of them (40% of the total rating) are based on data retrieved from the Web of Science (for more information on this ranking the reader is referred to Liu & Cheng, 2005; van Raan, 2005; Docampo 2011; Aguillo et al., 2010). Since 2007 it offers five rankings by field and since 2009, five rankings by subject.
>
> 2) *QS Ranking* (http://www.topuniversities.com/). The first edition of this ranking was launched in 2004. Until 2009 it was produced in partnership with the Times Higher Education, however, since then each company develops its own ranking (for more information on this ranking the reader is referred to Aguillo et al., 2010; Usher & Savino, 2007). 20% of the total rating assigned to each university is based on data retrieved from the database Scopus. It offers along with the global league table, 29 rankings by discipline classified into five major fields.
>
> 3) *NTU Ranking* (http://nturanking.lis.ntu.edu.tw). This ranking was first launched in 2007. It aims at measuring solely the quality of universities' research. It is based on 8 indicators all of them supported by bibliometric data from the Web of Science and the Thomson Reuters Essential Science Indicators (for more information on this ranking the reader is referred to e.g., Aguillo et al., 2010). Along with the global table league, it offers rankings by field and subject in a similar structure to that of the Shanghai Ranking. In this case, it offers 6 rankings by field and 14 rankings by subject.
>
> 4) *Leiden Ranking* (http://leidenranking.com). The first version of the Leiden Ranking was published in 2008[3]. However it was discontinued and, despite a 2010 edition was announced it is no longer available. In 2012 they resumed their activity and is now updated on an annual basis. Its methodology, shortcomings and potential use are discussed by Waltman et al. (2013). In its latest edition, it includes for the first time, rankings by five broad fields. These fields are constructed based on aggregations of the Web of Science subject categories. In the case a journal is assigned to several fields, its publications are assigned fractionally. The assignment of subject categories is available at the Methodology section of their website.

---

[3] Available at http://www.cwts.nl/ranking/LeidenRankingWebSite.html

Table 4. Main characteristics of the Shanghai Ranking, QS Ranking, NTU Ranking and Leiden Ranking

| Ranking | Shanghai Ranking | QS Ranking | NTU Ranking | Leiden Ranking |
|---|---|---|---|---|
| Launch year | 2003 | 2004* | 2007 | 2008** |
| 1st edition with fields | 2007 | 2009 | 2007 | 2013 |
| No. of fields | 5 | 5 | 6 | 5 |
| No. of subjects | 5 | 29 | 14 | 0 |
| Total universities | 500 | +701 | 500 | 500 |
| Type of data | Bibliometric and reputational | Bibliometric, surveys and manpower | Bibliometric | Bibliometric |
| Bibliometric data sources | Web of Science | Scopus | Web of Science | Web of Science |
| Ranking focus | Research & Teaching | Research, Teaching & Innovation | Research | Research |
| Weight of research performance indicators | 90% | 40% | 100% | 100% |

\* It offered a joint ranking in collaboration with the Times Higher Education Suplement, since 2009 it offers an independent ranking
\*\* Although its first edition dates back to 2008, it has not been published regularly since 2012. Since then it is published annually.

At this point it is important to note that the THE Rankings are not included in this study. Although they meet the criteria we do not include them for technical reasons. Only four Spanish universities are included in three of their six rankings by fields. Such a low presence does not allow its analysis and comparison with the national ranking. Also the Scimago Institutions Rankings are missing from this analysis. This is because they do not provide rankings by fields in their last edition.

*3.2 Concordance between international and national rankings and levels of agreement*

In order to establish fair comparisons and provide a global picture of the state of Spanish universities using national and international rankings, we first need to ensure that the classification of fields of national and international rankings is somehow similar and therefore, compatible. For this, we would need to analyze the way these fields are constructed for the four rankings used in this study and determine to which grade the methodology employed by each of them allows fair comparisons. As mentioned before, the I-UGR Rankings construct fields and disciplines by aggregating the Thomson Reuters subject categories. The Leiden Ranking and the NTU Ranking use the same approach, and the construction of fields and subjects is declared at their website. However, this does not occur for the other two rankings, which do not declare the methodology employed for establishing such fields. This lack of transparency is a shortcoming that must be taken into account when using these rankings for research policy.

We analyzed the fields and subjects of the selected international rankings and we established the homologous field or discipline according to the I-UGR Rankings. In Tables 4-7 we show the matching of fields per ranking. In general terms, we observe that it is possible to match most of the fields between the four international rankings selected and the I-UGR Rankings, although some exceptions are noted. The areas misrepresented in the I-UGR Rankings were Mechanical Engineering (QS Ranking and NTU Ranking), Law (QS Ranking) and all of the areas considered of the Arts & Humanities fields by the QS Ranking. This is due to the way the I-UGR Rankings are constructed, as they rely on the JCR and these lack journal rankings for these fields. Also, we observe that some fields of the international rankings (i.e., the Shanghai Ranking and the field of Social Science) include more than one of the fields included in the I-UGR Rankings. Finally, the classification of fields and subfields does not always match between rankings. Although this issue has no relevance for the purposes of this analysis, we must point out that subjects considered as major areas in one ranking are considered in the other as subfields or disciplines.

The four selected rankings included a total of 33 Spanish universities dispersed in 51 different fields and subfields. In Tables 4-7 we show the levels of agreement between international and national rankings according to the assignment of areas. For each area we calculate the Spearman coefficient to analyze the consistency between both rankings and the number of universities included in international rankings which take up the top positions of the national ranking. That is, if 6 Spanish universities are included in an international ranking but only two occupy positions between 1 and 6, the coincidence will be 2/6.

**Table 5.** Matching of fields and disciplines between the Shanghai Ranking and the I-UGR Rankings

| SHANGHAI RANKING | I-UGR RANKINGS | RHO | A |
|---|---|---|---|
| Natural Sciences & Mathematics | Mathematics / Physics / Chemistry | -0.866; 0; -0.866 | 0/3; 3/3; 2/3 |
| Engineering/Technology & Computer Sciences | Engineering / Information & Communication Technology | * | 1/3; 3/3 |
| Life & Agricultural Sciences | Agricultural Sciences / BiologicalSciences | * | 0/2; 1/2 |
| Clinical Medicine & Pharmacy | Medicine & Pharmacy | * | 2/2 |
| Social Science | Other Social Sciences / Psychology & Education / Economics, Finance & Business | * | 0/2; 0/2; 1/2 |
| Mathematics | Mathematics | -0.817 | 6/9 |
| Physics | Physics | 0.179 | 6/7 |
| Chemistry | Chemistry | 0.523 | 7/9 |
| Computer Science | Computer Science | 0.677 | 6/9 |
| Economics & Business | Economics, Finance & Business | 0.000 | 2/3 |

**Note:** Rho indicates the Spearman's coefficient. A indicates the level of agreement between rankings, that is, the number of universities present in both rankings.
*Insufficient values to calculate the indicator

The highest coincidence of universities between those present in the international rankings and top positions in the national ranking can be found in the NTU Ranking (Table 7), with 77.90% of the universities coinciding in both rankings. This ranking is followed by the Shanghai Ranking (Table 4) with 75.51% of the universities and the Leiden Ranking (Table 6) with 72.60%. The ranking with a lower percentage of coincidence is the QS Ranking (Table 5) with 56.49% of the universities present in this ranking reaching top positions in the national ranking.

Analyzing the fields we find the following disciplinary concordance:

- The Shanghai Ranking is the less consistent with the I-UGR Rankings showing positive low correlation in two fields (Chemistry and Computer Science).

- The NTU Ranking shows correlations above 0.7 in 9 out of 17 fields. The three fields with the highest correlations can be found between the NTU Ranking and the I-UGR Rankings and these are Physics (0,952), Chemistry (0.945) and Biological Sciences (0,886).

- The QS Ranking shows correlations above 0.7 in 8 out of 23. The fields of Biological Sciences (0.866) and Life Sciences & Medicine (0.882 with Medicine & Pharmacy) are the fields with a higher correlation.

- The Leiden Ranking only shows a correlation above 0.7 in one field, Natural Sciences & Engineering, with the field of Chemistry in the national ranking.

**Table 6.** Matching of fields and disciplines between the QS Ranking and the I-UGR Rankings

| | QS RANKING | I-UGR RANKINGS | RHO | A |
|---|---|---|---|---|
| | Arts & Humanities | | | |
| | Engineering & Technology | Engineering | 0.343 | 10/12 |
| | LifeSciences & Medicine | Biological Sciences / Medicine & Pharmacy | 0.609;0.882 | 9/11; 10/11 |
| | Natural Sciences | Mathematics / Physics / Chemistry | -0.518; 0.773; 0.700 | 10/11; 9/11; 8/11 |
| | Social Sciences & Management | Other Social Sciences/Psychology & Education/Economics, Finance & Business | 0.545; 0.155; 0.482 | 8/11; 7/11; 7/11 |
| Arts & Humanities | Philosophy | | | |
| | Modern Languages | | | |
| | Geography | Geography & City Planning | 0.775 | 2/4 |
| | History | | | |
| | Linguistics | | | |
| | English Language & Literature | | | |
| Engineering & Technology | Computer Science & Information Systems | Computer Science | 0.707 | 2/5 |
| | Chemical Engineering | Chemical Engineering | 0.463 | 3/6 |
| | Civil Engineering | Civil Engineering | 0.500 | 1/3 |
| | Electrical Engineering | Electric & Electronic Engineering | 0.154 | 3/6 |
| | Mechanical Engineering | | * | |
| Life Sciences & Medicine | Medicine | Medicine | | 1/2 |
| | Biological Sciences | Biological Sciences | 0.866 | 3/3 |
| | Psychology | Psychology | 0.414 | 4/6 |
| | Pharmacy & Pharmacology | Pharmacy & Toxicology | 0.507 | 5/6 |
| | Agriculture & Forestry | Agriculture | 0.461 | 4/10 |
| Natural Sciences | Physics & Astronomy | Physics | 0.775 | 3/4 |
| | Mathematics | Mathematics | -0671 | 2/5 |
| | Environmental Sciences | Earth & Environmental Sciences | 0.632 | 2/4 |
| | Earth & Marine Sciences | Earth & Environmental Sciences | * | 1/2 |
| | Chemistry | Chemistry | 0.447 | 2/4 |
| | Materials Science | Materials Science | * | 2/3 |

| | Statistics & Operational Research | Statistics | 0.612 | 6/10 |
|---|---|---|---|---|
| | Sociology | Sociology | -0.289 | 3/5 |
| | Politics & International Studies | Political Science | * | 0/1 |
| Social Sciences & Management | Law | | | |
| | Economics & Econometrics | Economics | 0.754 | 4/6 |
| | Account & Finance | Business | 0.775 | 3/4 |
| | Communication & Media | Communication | 0.289 | 0/5 |
| | Education | Education | 0.158 | 3/5 |

Note**:** Rho indicates the Spearman's coefficient. A indicates the level of agreement, that is, the number of universities present in both rankings.
*Insufficient values to calculate the indicator

**Table 7.** Matching of fields and disciplines between the Leiden Ranking and the I-UGR Rankings

| LEIDEN RANKING | I-UGR RANKINGS | RHO | A |
|---|---|---|---|
| Biomedical & Health Sciences | Medicine & Pharmacy | 0.518 | 10/15 |
| Life & Earth Sciences | Biological Sciences / Earth & Environmental Sciences | 0.600; 0.436 | 11/15; 9/15 |
| Mathematics & Computer Science | Mathematics / Information & Communication Technology | 0.307; -0.036 | 12/15; 9/15 |
| Natural Sciences & Engineering | Engineering / Mathematics / Physics / Chemistry | 0.350; 0.264; 0.496; 0.736 | 12/15; 12/15; 12/15; 11/15 |
| Social Sciences & Humanities | Other Social Sciences / Psychology & Education / Economics... | 0.121; 0.115; 0.220 | 8/13; 8/13; 8/13 |

Note**:** Rho indicates the Spearman's coefficient. A indicates the level of agreement.*Insufficient values to calculate the indicator

**Table 8.** Matching of fields and disciplines between the NTU Ranking and the I-UGR Rankings

| NTU RANKING | I-UGR RANKINGS | RHO | A |
|---|---|---|---|
| Agriculture | Agriculture | 0.406 | 5/10 |
| Clinical Medicine | Medicine | * | 2/2 |
| Engineering | Engineering | 0.418 | 9/11 |
| Life Sciences | Biological Sciences | 0.886 | 5/6 |
| Natural Sciences | Mathematics / Physics / Chemistry & Chemical Engineering | 0.127; 0.879; 0.588 | 6/10; 8/10; 8/10 |
| Social Sciences | Other Social Sciences / Psychology & Education / Economics… | 0.600; 0.000; 0.400 | 2/4; 2/4; 2/4 |
| Agricultural Sciences | Agricultural Sciences | 0.440 | 18/23 |
| Environment/Ecology | Earth & Environmental Sciences | * | 0/0 |
| Plant & Animal Science | Biological Sciences | 0.552 | 5/10 |
| Computer Science | Computer Science | 0.812 | 13/16 |
| Chemical Engineering | Chemical Engineering | 0.846 | 8/12 |
| Civil Engineering | Civil Engineering | 0.202 | 10/12 |
| Electrical Engineering | Electrical & Electronic Engineering | 0.755 | 8/11 |
| Mechanical Engineering | | | |
| Materials Science | Materials Science | 0.757 | 5/7 |
| Pharmacology | Pharmacy & Toxicology | 0.300 | 5/5 |
| Chemistry | Chemistry | 0.945 | 14/15 |
| Geosciences | Geosciences | 0.847 | 6/7 |
| Mathematics | Mathematics | 0.524 | 11/12 |
| Physics | Physics | 0.952 | 7/8 |

**Note:** Rho indicates the Spearman's coefficient. A indicates the level of agreement.*Insufficient values to calculate the indicator

If we focus on the disciplinary differences, the coincidence is especially relevant for the fields and subjects of Biomedicine, Life Sciences and Natural Sciences. This does not occur in the Social Sciences where the only exception noted is Economics.

In the case of rankings, in general terms, we can point out the following lessons learned:

- The NTU Ranking is the one which seems to be more consistent with the I-UGR Rankings. This is not surprising as it measures solely the research dimension and is fully based on the Web of Science, as it occurs with the I-UGR Rankings. Also, the confection of the fields and subfields is similar as both rankings aggregate subject categories to construct the fields, while in the other two cases this is not explained.

- Though this could be expected also with the Leiden Ranking, it does not occur mainly due to two reasons. Firstly, the Leiden Ranking is based on a fixed set formed by the 500 most productive universities worldwide and in all areas. This means that universities with a lower overall output but significantly outstanding in certain fields are not included in the rankings by fields. Secondly, the indicator used in this study for sorting the universities is the proportion of top 10% publications. This indicator is based on the qualitative dimension of the research outcome. As the I-UGR Rankings employ the $IFQ^2A$ Index which contemplates both, the qualitative and quantitative dimension, this may affect the correlation between rankings.

- Another issue which affects this in the other two ranking (Shanghai Ranking and QS Ranking) has to do with the way results are presented, as they only show the intervals in which each university is positioned after they surpass certain threshold. Although the QS Ranking provides the rating of each university, allowing the user to rank universities, this does not occur with the Shanghai Ranking.

## 4. Concluding remarks and lessons learned

In this paper we explore the possibility of using national rankings to complement international rankings, as the latter usually offer a poor representation of national university systems (no more than 25% of the system in the Spanish case). We insist on the importance of rankings by fields (García et al., 2012) as these do not neglect universities' disciplinary focus and offer a complete picture of universities' research performance. This perspective follows the recent trend on evaluative bibliometrics for 'opening up' these tools in order to offer, rather than a narrow and simplistic solution, a range of different outputs that can better serve research policy makers to make the right decisions considering their specific aims and different scenarios (Rafols et al., 2012).

We use Spain as a study case and we introduce the I-UGR Rankings for Spanish universities. This ranking uses the $IFQ^2A$ Index, an indicator which measures the qualitative as well as the quantitative dimension of research (Torres-Salinas et al., 2011c). From this analysis we conclude that national rankings can complement international rankings in order to provide a complete picture of university systems despite the methodological differences aroused from the comparisons by fields. However, we must stress the importance of acknowledging such methodological differences to better interpret them. Such differences are mainly derived from the construction of fields and subfields as well as the indicator selected for ranking universities.

Our conclusion is clear as to the importance and complement that represent the national rankings to address a comprehensive analysis of the university system of a country. The joint analysis of both types of rankings will provide a complete snapshot of the universities and their scientific strengths.

These results show different levels of concordance which are affected not only by methodological issues but also by the way these fields are constructed and the difficulties implied in this process which affected differently each scientific domain. Despite this, it is possible to use both (national and international rankings) and combine the information provided in a research policy context.

### Acknowledgments
Thanks are due to the two anonymous referees for their helpful comments. Nicolás Robinson-García is currently supported by a FPU grant from the Ministerio de Educación y Ciencia of the Spanish Government.